\documentclass{sig-alternate-10pt}
\usepackage{endnotes}
\usepackage{epsfig}
\usepackage{amssymb}
\usepackage{amsmath}
\usepackage{algorithm}
\usepackage{algorithmic}
\usepackage{amsfonts}
\usepackage{graphicx}

\usepackage{amsfonts,amsmath,amstext}

\begin{document}

\date{}

\title{Divide and Conquer: Partitioning Online Social Networks}
\numberofauthors{3} 
\author{
\alignauthor
Josep M. Pujol\\
       \affaddr{Telefonica Research}\\
       \email{\normalsize jmps@tid.es}
\alignauthor
Vijay Erramilli \\
       \affaddr{Telefonica Research}\\
       \email{ \normalsize vijay@tid.es}
\and
\alignauthor Pablo Rodriguez\\
      \affaddr{Telefonica Research}\\
      \email{\normalsize pablorr@tid.es}
}

\maketitle
\vspace{-10mm}
\begin{abstract}
\label{sec:abs}
Online Social Networks (OSNs) have exploded in terms of scale and scope
over the last few years. The unprecedented growth 
of these networks present challenges in terms of system design and
maintenance. One way to cope with this is by partitioning 
such large networks and assigning these partitions to different
machines. However, social networks possess unique properties that 
make the partitioning problem non-trivial.  
The main contribution of this paper is to understand
different properties of social networks and how these properties 
can guide the choice of a partitioning algorithm. Using large scale 
measurements representing real OSNs, we first characterize different properties of 
social networks, and then we evaluate qualitatively different 
partitioning methods that cover the design space. We expose different 
trade-offs involved and understand them in light of properties of social networks.
We show that a judicious choice of a partitioning scheme can help improve 
performance.   

\end{abstract} 



\vspace{-3mm}
\section{Introduction} 
\label{sec:intro}
The last few years have seen wide-scale proliferation of
Online social networks (OSNs). According to a recent
study, OSNs have become more popular than email \cite{url:soc}
and popular OSNs like Facebook, MySpace, Orkut etc.
have tens of millions of active users, with more users
being added by the day. Twitter, for instance, grew
a remarkable 1382\% in one month(Feb-Mar ’09) \cite{url:nielsen}.
The growth of OSNs pose unique challenges in
terms of scaling, management and maintenance of these
networks \cite{url:linkedin}. In addition, online services are moving
towards an increasingly distributed cloud paradigm \cite{echos,cloud-ccr,church}, 
that bring forth their own challenges since smaller and 
geographically spread data centers \cite{cloud-ccr} 
will increase network costs.

One obvious solution to deal with scaling issues would be partition
the social network graph and assign partitions to different servers.  
However, given the unique properties of
social network graphs (presence of clusters or communities,
skewed degree distributions, correlations to geography),
it is not entirely clear \emph{how} to partition such
graphs, what \emph{properties} of the underlying social network
graph need to be taken into account and what are
the \emph{trade-offs} involved in partitioning.
Understanding the design space and the trade-offs can help in making
informed choices. 

The main contribution of this paper is to provide an understanding of
the problem of social network partitioning by relying on measurements 
taken from two large and popular social networks: Twitter and Orkut. 
We state the problem of social network partitioning 
along with the performance objectives, highlighting the
trade-offs involved. We then discuss the different characteristics 
of social network graphs that can impact design choices of partitioning 
algorithms. In particular, using the data we obtain for the Twitter network, 
we measure strong geographical locality as well as heterogenous traffic 
patterns between users. 

Based on the performance objectives and social network characteristics, we select
the following partitioning algorithms to evaluate against our data-sets. We first choose
METIS \cite{metis} that is drawn from traditional graph partitioning 
algorithms \cite{spectral,metis,dan}. This is an intuitive choice as these algorithms 
are designed to produce partitions while reducing traffic across partitions and 
balancing the number of nodes across partitions. Since it is widely known that 
social networks consist of `community' structure \cite{leskovec,newman-2006}, the next 
method we evaluate relies on extracting such communities, and these communities can 
be used as partitions. However certain issues with such algorithms, in particular  
arbitrary number of communities and skewed partition sizes, lead us to augment 
an existing community detection scheme to deal with these issues. We evaluate these algorithms against different performance metrics to expose various 
trade-offs, in particular the trade-off between balancing load across different partitions
and reducing traffic between partitions. 
After evaluating different algorithms against our datasets, 
we find that the augmented community detection that we devised does well in terms of 
balancing the trade-offs involved and we interpret 
these results in light of properties that social networks possess. 

\vspace{-4mm}
\section{Related Work} 
\label{sec:relwork}
Graph partitioning has been proposed to deal with scaling issues of social 
network systems \cite{url:linkedin, url:facebook}, as well as handling large 
data-sets \cite{botgraph}. However, to the best of our knowledge, there is no
study of different properties of social network graphs, how these properties can
impact the possible different choices of partitioning algorithms and 
how different partitioning algorithms perform on real data, along with quantifying 
the different tradeoffs involved. We aim to address these concerns in this paper.

There has been a concerted effort to characterize and understand Online Social Networks 
over the last few years \cite{mislove-wosn, moon,walter-wosn,twitter,mislove}, including
flow of information \cite{kdd-watts}, existence of social communities \cite{leskovec} as 
well as evolution of such networks \cite{mislove-wosn,kdd-jon}. 
As such, we are more focused on understanding the aspects of social networks
that can impact partitioning. Towards that end, we rely on certain 
results (like existence of social communities \cite{leskovec,newman03}) that have been 
reported in the past and also report existence of certain properties (like geographical
locality and heterogenous traffic distribution) in one network we study; Twitter 
\cite{twitter} that can better inform system architects interested in social networks, as well as design of partitioning schemes.

In order to understand the trade-offs involved, we rely on
different partitioning schemes that span a range of design choices. 
We first focus on a method drawn from classical
graph partitioning algorithms \cite{spectral,metis, dan,teng} 
that rely on finding minimum edge cuts to
seperate the graph into roughly equal-sized clusters. We also draw from work done that 
aims to extract `communities' that are socially relevant \cite{newman-2006,blondel-2008}. 
However these algorithms do not give a desired size of communities, nor do they give
equal sized partitions. We present a scheme in this paper that yields
a desired number of communities, of equal size.

\vspace{-4mm}
\section{Methodology} 
\label{sec:method}
In this section, we pose the problem of social network partitioning, along with the 
objectives. We then explore properties of social network graphs that could impact 
the design and choice of different partitioning schemes. We end this section with 
a brief description of some of the methods we explore in this study. 
We first describe the data-sets that we collect and use. 

%
%

\begin{figure*}[t]
\centering
\begin{tabular}[1]{ccc}
\includegraphics[width=2.2in]{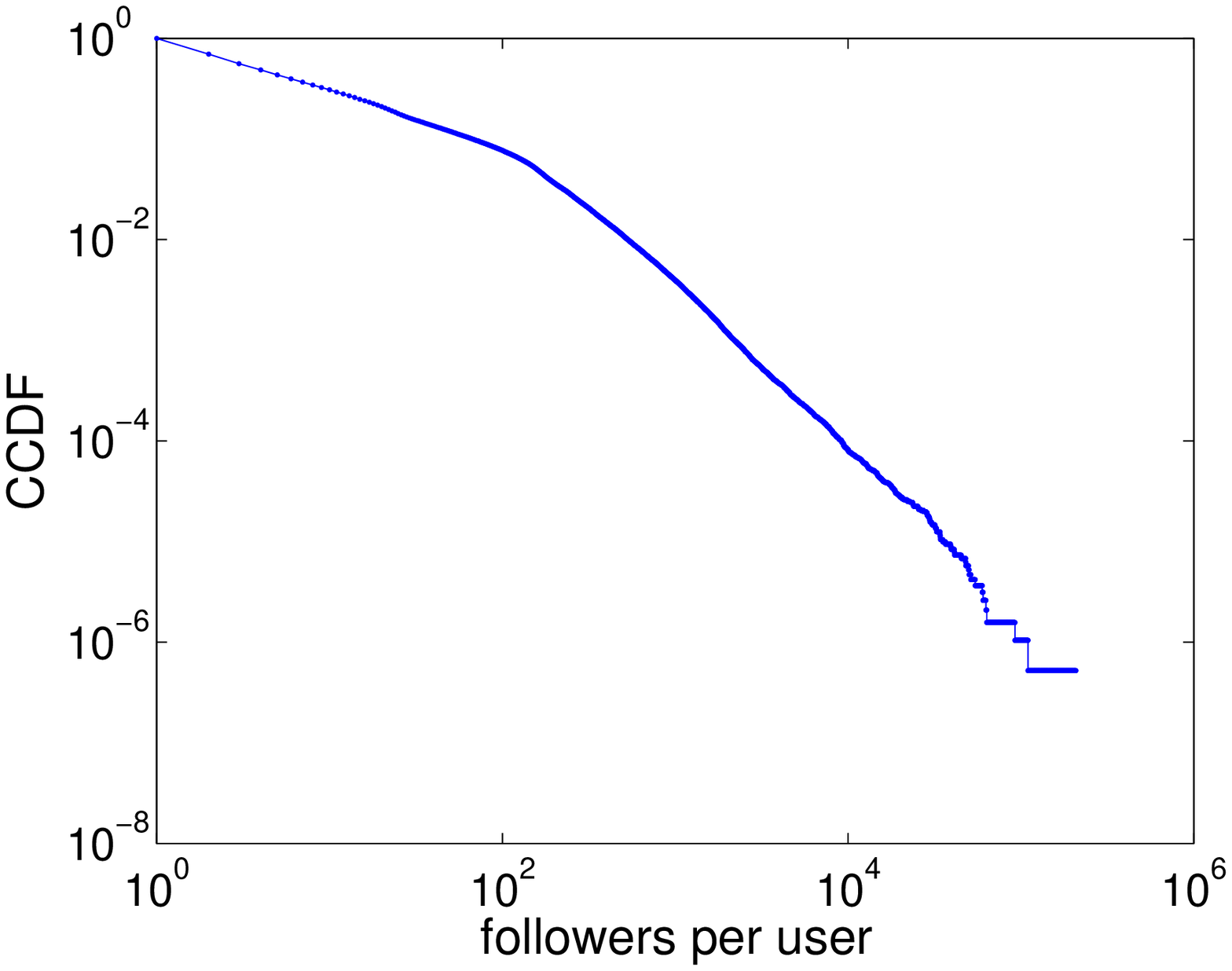}
&
\includegraphics[width=2.2in]{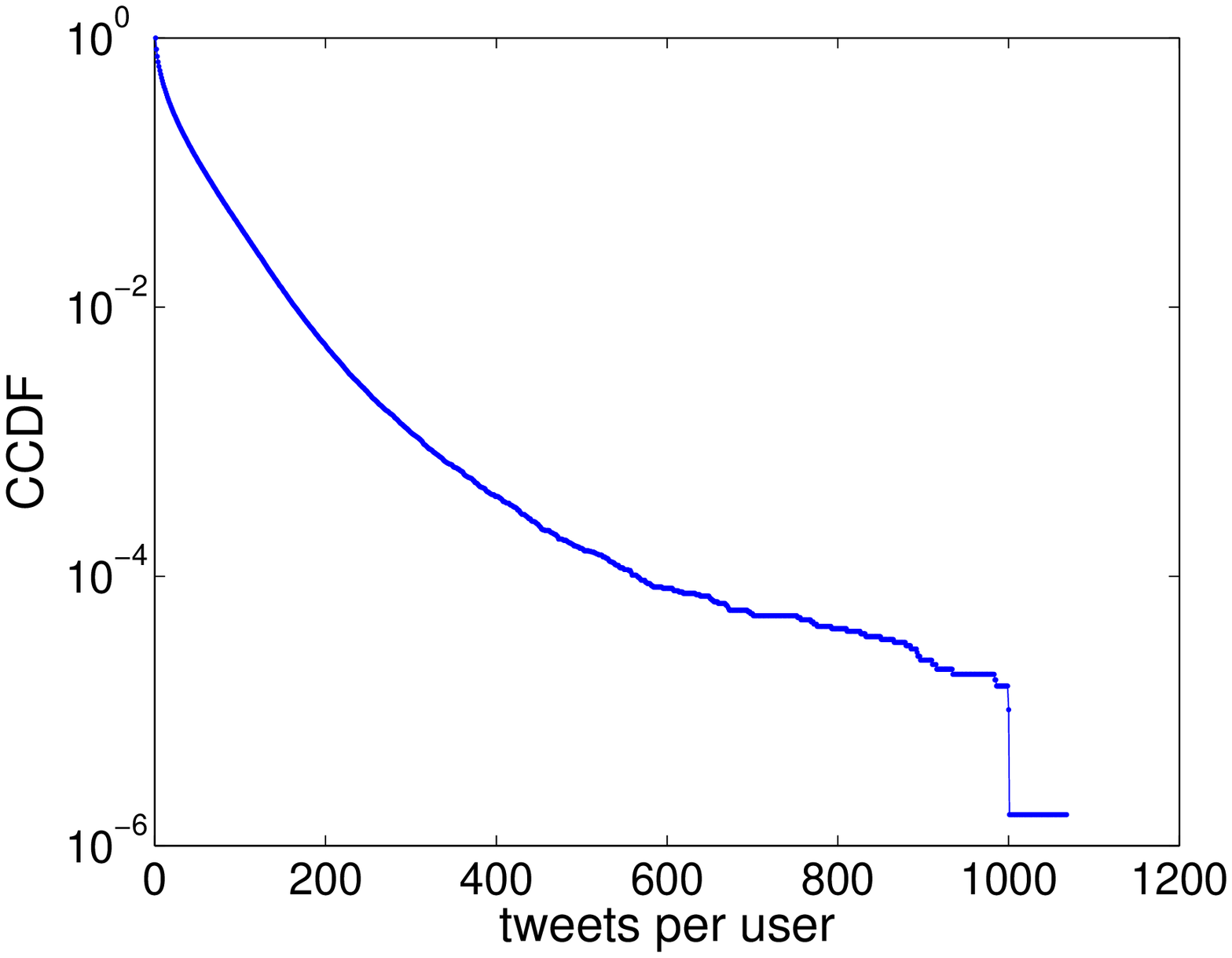}
&
\includegraphics[width=2.2in]{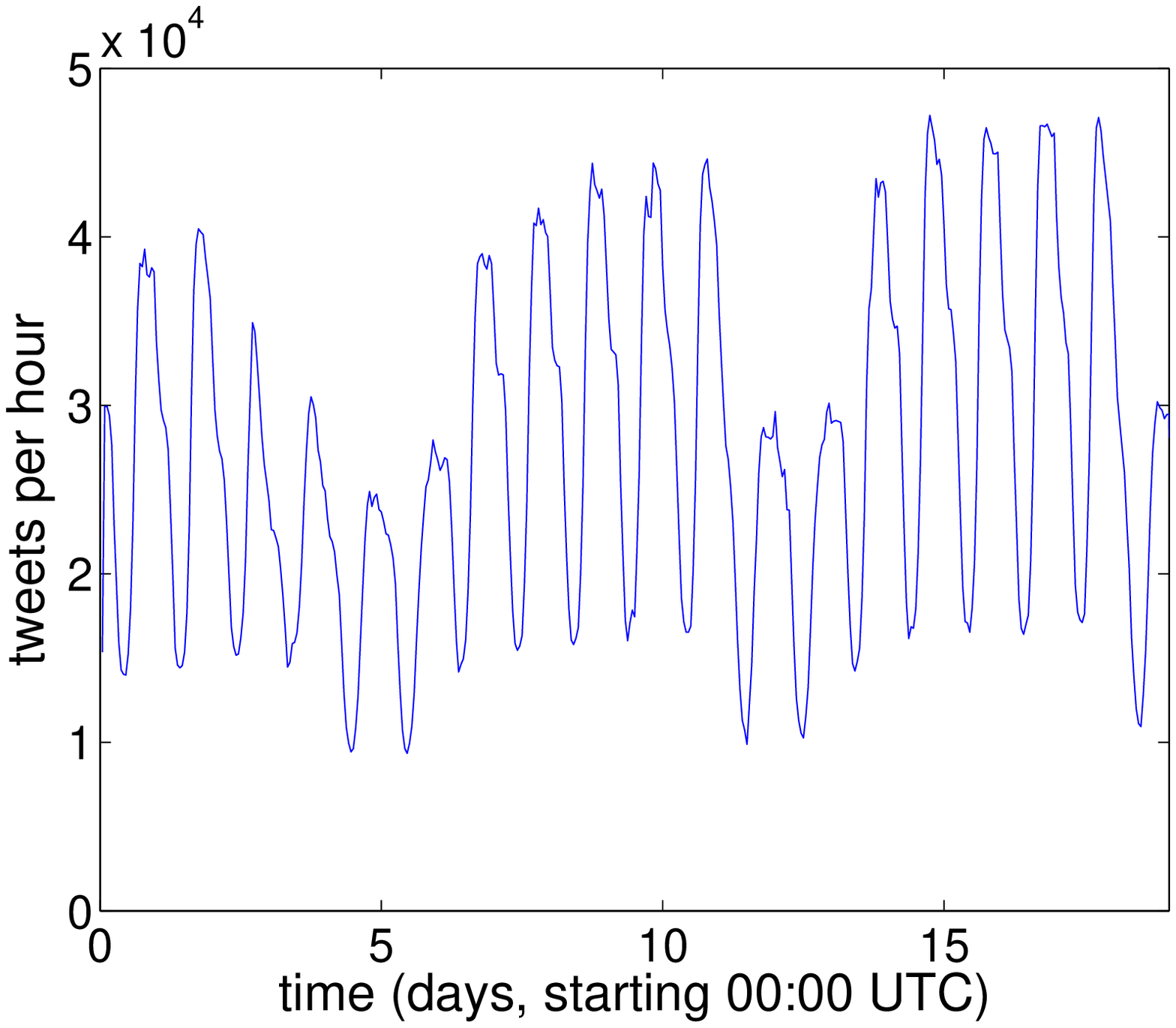}
\\
(a) & (b) & (c)
\end{tabular}
\caption{Twitter Characteristics (a) CDF of Followers
(b) CDF of 'Tweets'/Messages per user (c) 'Tweets' per hour}
\label{fig:twit_traffic}
\vspace{-5mm}
\end{figure*}
\vspace{-2mm}

\subsection{Data} 
In order to carefully understand the different characteristics of social 
network graphs that can impact partitioning, we rely on data. There exist
data-sets of large social networks that are publicly available \cite{mislove}. 
However, we were looking for data-sets that included location information, as well
as information about traffic that is exchanged between users. Since we are not aware 
of any publicly available data-sets, we collected our own data-set by 
crawling Twitter (http://www.twitter.com). Twitter is a 
microblogging site that has become very popular of late.  
More details about Twitter, along with information on how links are formed, 
can be found in \cite{twitter}. We collected data from Twitter between Nov 25 - Dec 4, 2008
and collected information comprising $2,408,534$ nodes and $38,381,566$ edges (although
Twitter has directed edges, we report total edges and use edges as undirected 
unless otherwise stated). 

In addition, we also collect location information, as self-identified by users. 
This can be in the form of free-text or by latitude/longitude. 
In order to glean locations, we had to filter the free-text for junk information.   
After basic preprocessing we obtained 187K different locations for the 996K users 
where the location text was meaningful. 
We then used 
the Yahoo! Maps Web Services\footnote{http://developer.yahoo.com/maps/rest/V1/geocode.html} 
that offers the Yahoo Maps functionality as an API to standardize locations.
We could test the 187K different locations to obtain the country, and where available, 
the state and the city. 
We finally obtained the standardized locations for 691K users.

We also collected traffic on Twitter in the form of `tweets' (total 12M tweets) 
between the 2.4M users by using the Twitter API\footnote{http://apiwiki.twitter.com}. 
From the tweets, we mined the following relevant 
fields: {\em tweet\_id}, {\em timestamp}, {\em user\_id}, {\em location} and 
{\em content}. This information allows us to 
get traffic information for every user and
how this traffic is distributed across users (this gives us volume of social conversations).
Approximately 25\% of the population (587K) generated at least one tweet, 
for the rest of the users the Twitter API did not return tweets in the 19-days period 
under examination.  
Fig.~\ref{fig:twit_traffic}(a) and (b) show the skewed distribution of the number of links
per user as well as the volume of traffic sent per user. 
We intend to make this data public.

We also use graphs representing other social networks, including Orkut, that can
be taken to be closer to an actual `social' network, as edges in such a graph 
will probably represent actual social relationships. This dataset consists of 
$3,072,441$ nodes and $223,534,301$ edges and this data was collected 
between Oct 3 - Nov 11, 2006. More information can be found in \cite{mislove}.
We present results from Twitter and Orkut in this paper for space reasons, although 
our analysis on other graphs \cite{mislove} have borne similar results. 

\subsection{Social Network Partitioning}
We represent a social network as a graph $G~=~(V,E)$, where the edges are undirected. 
The problem then is to partition $V$ into $k$ subsets or clusters, 
$(V_1, V_2,..,V_k)$, $k>1$ such that 
$V_i \cap V_j = \emptyset$ when $i \neq j$ and $\cup_i V_i = V$. The objective function 
to decide how to obtain these $k$ partitions differ. 

\textbf{Performance Objectives} The first objective function 
for a partitioning scheme would be to reduce the edges that traverse 
between partitions. We refer to these edges as external or \emph{inter}-edges. 
This metric represents the amount of traffic that could traverse 
between partitions and reducing this metric therefore, 
has a direct bearing on reducing bandwidth costs. Bandwidth costs 
can be a bigger concern if a social network graph is hosted across geographically 
distributed data-centers \cite{echos}.

The second objective function is related to keeping utilization high and balanced 
across servers hosting different partitions. Ideally, 
all partitions should be of \emph{equal} size, leading to balanced utilization 
across machines. It has been noted that given the costs of server installation 
and limited lifetime, keeping servers on and running optimizes work 
per investment dollar \cite{cloud-ccr}. By assigning nodes to different servers
(equivalent to assigning partitions) in an informed manner 
can help achieve this goal.

Balancing load across servers and 
reducing inter-server traffic are often at odds with each other and 
hence there is a need to study the design space of partitioning schemes to
characterize the tradeoffs involved. In addition, social networks have 
slightly different underlying characteristics compared to regular communication networks
that can be exploited for the design of partitioning schemes. 
We discuss these characteristics in the next section.

\begin{table*}[t]
\begin{minipage}[b]{0.5\linewidth}\centering
\begin{tabular}{|l|l|l|l|l|| }
 
\hline Country & Size (\%) & \% of inlinks & \% of outlinks  \\ \hline
 US & 60.2  & 79.9 & 81.0  \\
GB  & 6.3 & 32.9 & 31.8 \\
CA & 4.02 & 26.7 & 25.2 \\
BR  & 3.0 & 65 & 62.2 \\
JP  & 2.9 & 80  & 83 \\ \hline
 \end{tabular}
   \vspace{-4mm}
\end{minipage}
\begin{minipage}[b]{0.5\linewidth}
\centering
\begin{tabular}{|l|l|l|l|l|}
 
\hline State & Size (\%) & \% of inlinks & \% of outlinks  \\ \hline
CA & 16.9  & 32 & 37  \\
NY  & 8.3 & 22 & 25 \\
TX & 7.2 & 28 & 28 \\
FL  & 4.6 & 19 & 17.6 \\
IL  & 4.1 & 20  & 20 \\ \hline
 \end{tabular}
\vspace{-4mm}
  \end{minipage}
\caption{Presence of Strong GeoLocality (Countries and US States): Twitter } 
  \label{tab4}
  \end{table*}

\subsection{Properties of Social Networks} 
\label{subsec:prop}
Properties that can be relevant to the problem 
of partitioning are: 

\textbf{Structural Properties:} Social networks are known to have
skewed degree distributions, assortativity \cite{newman03}, 
small world properties \cite{watts99} as well as strong clustering or 
\emph{community} structure \cite{leskovec}. Partitioning by uncovering communities
in a social network can be intuitively appealing as volume of 
intra-community interaction is much higher than inter-community interaction. 

\textbf{Geo-Locality Properties:} Social networks have been shown to have strong correlations
to geography; the probability of new links forming is correlated to
distance \cite{lj-rbf}. We observe similar strong correlations to distance in the 
Twitter data-set. In Table~\ref{tab4}, we present locality results 
of the top $5$ countries and US states by size. We consider directed links. 
The second column shows the size distribution of
different countries; for instance US has $60.2\%$ of the total nodes. 
We also found the percentage of edges are 
closely correlated to the size of the network, for instance 66\% of the total inlink edges
and 65.01\% of the total outlink edges belong to the US. The next two columns represent the 
percentage of inlinks and outlinks that come from users of the same country. 
For instance, 80\% of inlinks in US come from nodes belonging to US, and 81\% of outlinks
in US also belong to nodes within US. If we compare these numbers to the 
relative sizes of the partitions, we realize that there is high \emph{geo-locality}; for a 
partition the percentage of links that are local is higher than what would be 
the case if connectivity is purely random; for instance in the case of US this would be 
$60.2\%$  
Factoring this locality property could yield performance 
benefits, specially if partitions are assigned to servers 
distributed geographically \cite{cloud-ccr,church}. Strong locality can thus reduce
inter-server traffic considerably. 

In addition, relying on semantic information like geographical location for partitioning 
can be appealing due to the simplicity as one does not need global information 
about the structure of the social network, and issues like \emph{churn} 
(users joining/leaving the social network) can be handled more effectively 
and efficiently. We do not propose a scheme based on geographical locality 
in this paper and leave it for future work. 
However, we note that the community structure embeds some information about locality \cite{lj-rbf}.

\textbf{Traffic Properties:}
Traffic on social networks primarily 
consists of messages, user-generated content (UGC) and status updates exchanged between
nodes. From our analysis of traffic on Twitter, we observed some 
similarities to traffic on IP networks like diurnal patterns (Fig.\ref{fig:twit_traffic}(c)).
However there are a few facets that makes traffic on social networks unique,
and hence worth paying special attention to. First of all, traffic in social networks 
is inherently \emph{local}; most traffic is confined to one-hop distance. Secondly, traffic 
in social networks can have a \emph{multiplicative} effect due to the broadcast nature of 
certain messaging protocols like status updates. To illustrate this point, consider 
Figs.\ref{fig:twit_traffic} (a, b). These figures show the skewed nature of the number of
followers (social contacts) a user in Twitter has, as well as the skewed nature of the
number of messages (tweets) sent per user. Given the broadcast nature of status updates
in Twitter, the 12M unique tweets, coupled with the skewed distribution of the number 
of followers, the actual number of tweets are in excess of 1.7B messages.

In addition, we also extracted \emph{conversations} between users using the content in 
tweets; we establish a link between user 
$i$ and $j$ $iif$ $i$-id appears in the content of at least one tweet of $j$ and 
vice-versa. We obtain 265K links between people who are maintaining a conversation, 
which is clearly less than the 38M links between people according to the followers SN.

This information captures the social network at a finer level than the social network 
given by a simple contact list; people have users in ones' contact list that they may 
not actually know \cite{moon}. Conversations imply a 
more deeper social relation, therefore they can be used to validate the 
accuracy of the partition algorithms in preserving such strong social relations. 

\begin{algorithm}
\caption{MO+}         
\label{algo1}
\begin{algorithmic}
\STATE {\bf Input:} $G = \left< V,E \right> $ Social network graph
\STATE {\bf Input:} $k$ Number of final partitions 
\STATE {\bf Input:} $\cal{F}$ Modularity optimization algorithm
\STATE {\bf Output:} $P$ Partition set, ${p_i}$ set of nodes of partition $i$, $|P|=k$  
\STATE $U \leftarrow \{ \{ 1..|V| \} \} $ 
\WHILE{$U \neq \emptyset$}
\STATE $G'\leftarrow \left < V \backslash v_i \notin U_1 , E \backslash e_{ij} ~|~ v_i \notin U_1 \vee  v_j \notin U_1 \right> $
\STATE $C \leftarrow {\cal F} \left( G' \right) $ --- partition algorithm
\STATE $C' \leftarrow C ~|~ |c_i| \geq |c_{i+1}|$ --- sort the best partition by size
\FOR{$c_i \in C'$}
\IF{$|c_i| > \frac{|V|}{k} $}
\STATE $U \leftarrow U \cup \{ c_i\}$
\ELSE 
\FOR{$p_j \in P$}
\IF{$c_i \neq \emptyset \wedge \left( |p_j|+|c_i| \leq \frac{|V|}{k} \right) $}
\STATE $p_j \leftarrow p_j \cup c_i$
\STATE $c_j \leftarrow \emptyset$
\ENDIF	
\ENDFOR
\ENDIF
\ENDFOR
\STATE $U \leftarrow U \backslash \{ U_1 \}$
\ENDWHILE
\end{algorithmic}
\end{algorithm}

\vspace{-3mm}
\subsection{Partitioning Methods}

In order to explore the design space for partitioning algorithms and quantify
the tradeoffs involved, we study the following methods:

\begin{figure*}[tbp]
\centering
\begin{tabular}[2]{cc}
\includegraphics[width=2.5in]{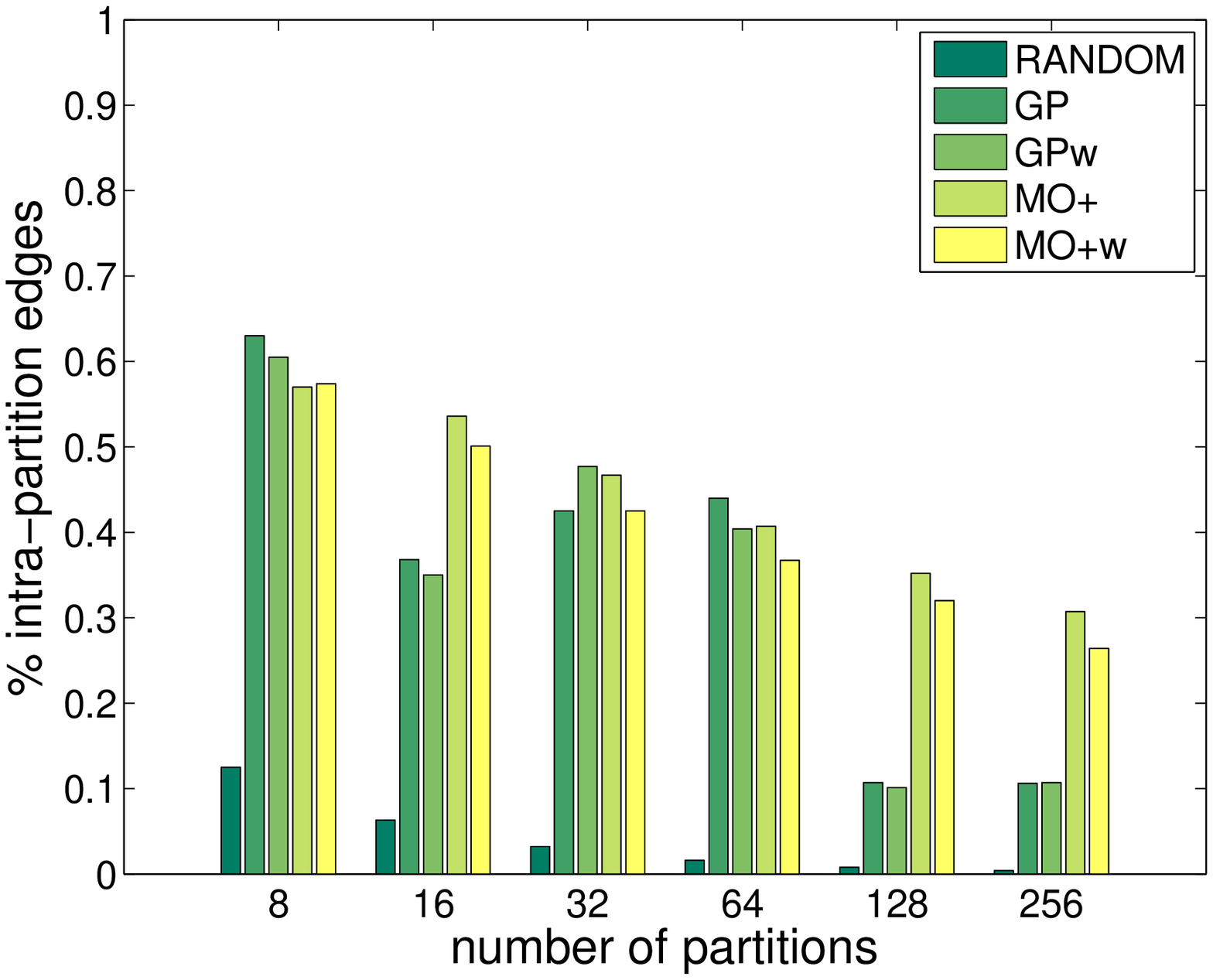}
&
\includegraphics[width=2.5in]{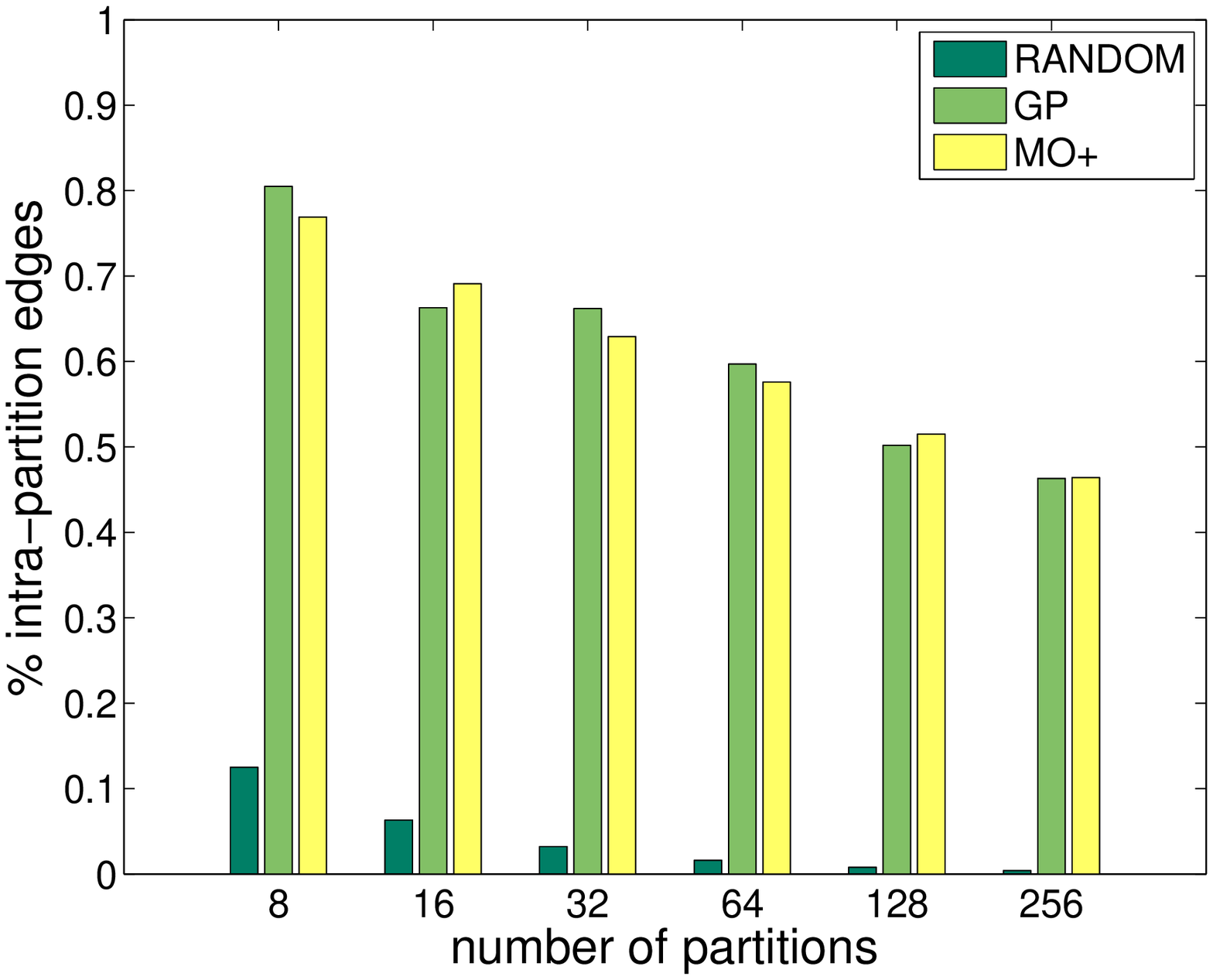}
\\
(a) & (b)
\end{tabular}
\caption{Partitioning Results: \%of Intra-Partition Edges vs \# of Partitions (a) Twitter, (b) Orkut }
\label{fig:graph_part}
\vspace{-4mm}
\end{figure*}

\textbf{Graph Partitioning (GP) Algorithms:} 
The objective function that these algorithms rely on is to reduce inter-cluster edges; 
reduce the number of edges incident on vertices belonging 
to different clusters. In addition, the partitions should be balanced; 
have similar sizes. Hence such algorithms are a natural candidate to study as the 
performance objectives for our problem are addressed by such algorithms.
Spectral partitioning algorithms rely on finding the 
minimum cut to ensure the minimum number of edges cross 
between partitions \cite{spectral}.  
Many different variants have been proposed and we rely on the Multi-way partitioning method,
also called METIS\cite{metis} that has been shown to produce very high 
quality clusters in a fast and efficient manner.

\textbf{Modularity Optimization (MO) Algorithms:} 
Social graphs with communities are intuitively different 
from random graphs in terms of structure \cite{newman03}. 
This difference can be captured via a metric called \emph{modularity} ($Q$) 
that is defined as the difference between the number of edges within 
communities and the expected number of such edges. 
Mathematically, $Q \sim A_{ij} - P_{ij}$, where $A_{ij}$ is the actual 
number of edges traversing two communities, and $P_{ij} \sim d{_i}*d{_j}$, 
where $d_i$ is degree of node $i$. The modularity metric 
is between $(0,1)$ and lower values signify a structure closer to random graphs, 
while higher values signify strong community structure. The algorithms developed 
for community detection therefore rely on finding partitions that maximize this 
metric. For the purposes of this study, the modularity optimization scheme we use is 
proposed in \cite{blondel-2008}, that suits our purpose and is capable of handling 
large graphs, very quickly. 
However, there are a couple of caveats: 
the communities can be unequal in size - this is closer to reality as real-life 
communities are seldom of equal size. 
The other problem is that there can be an arbitrary number of communities, as
the objective of such algorithms is to find the natural number of communities, 
rather than find a predefined number.  
When we ran the algorithm on our datasets, we got 2743 and 37 communities 
for Twitter (modularity=0.48) and Orkut (modularity=0.63) respectively, 
with highly skewed community sizes (size of largest community were 21\% and 25\% 
of the total number of nodes), making these scheme not amenable for using as-is.

\begin{figure*}[t]
\centering
\begin{tabular}[2]{cc}
\includegraphics[width=2.5in]{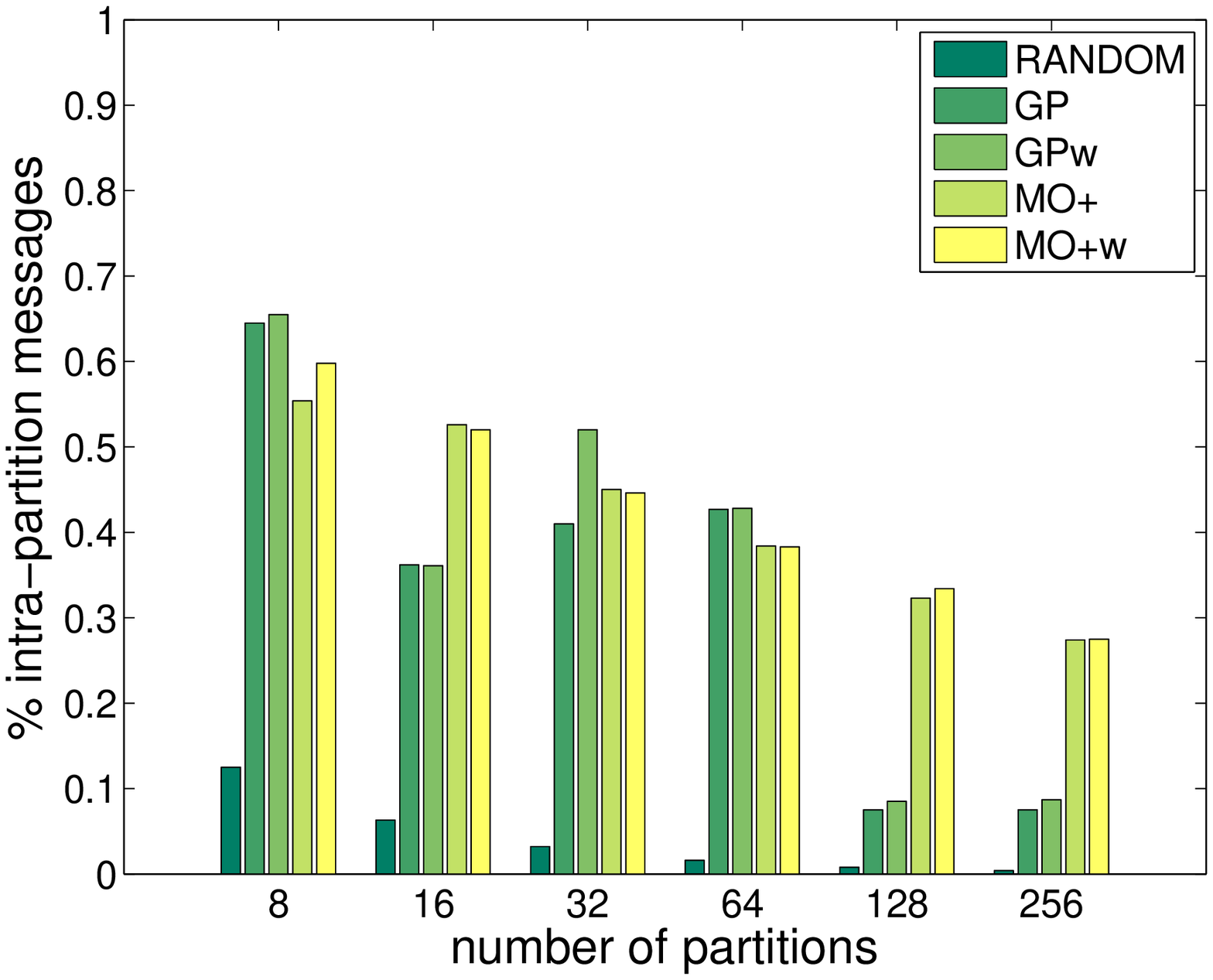}
&
\includegraphics[width=2.5in]{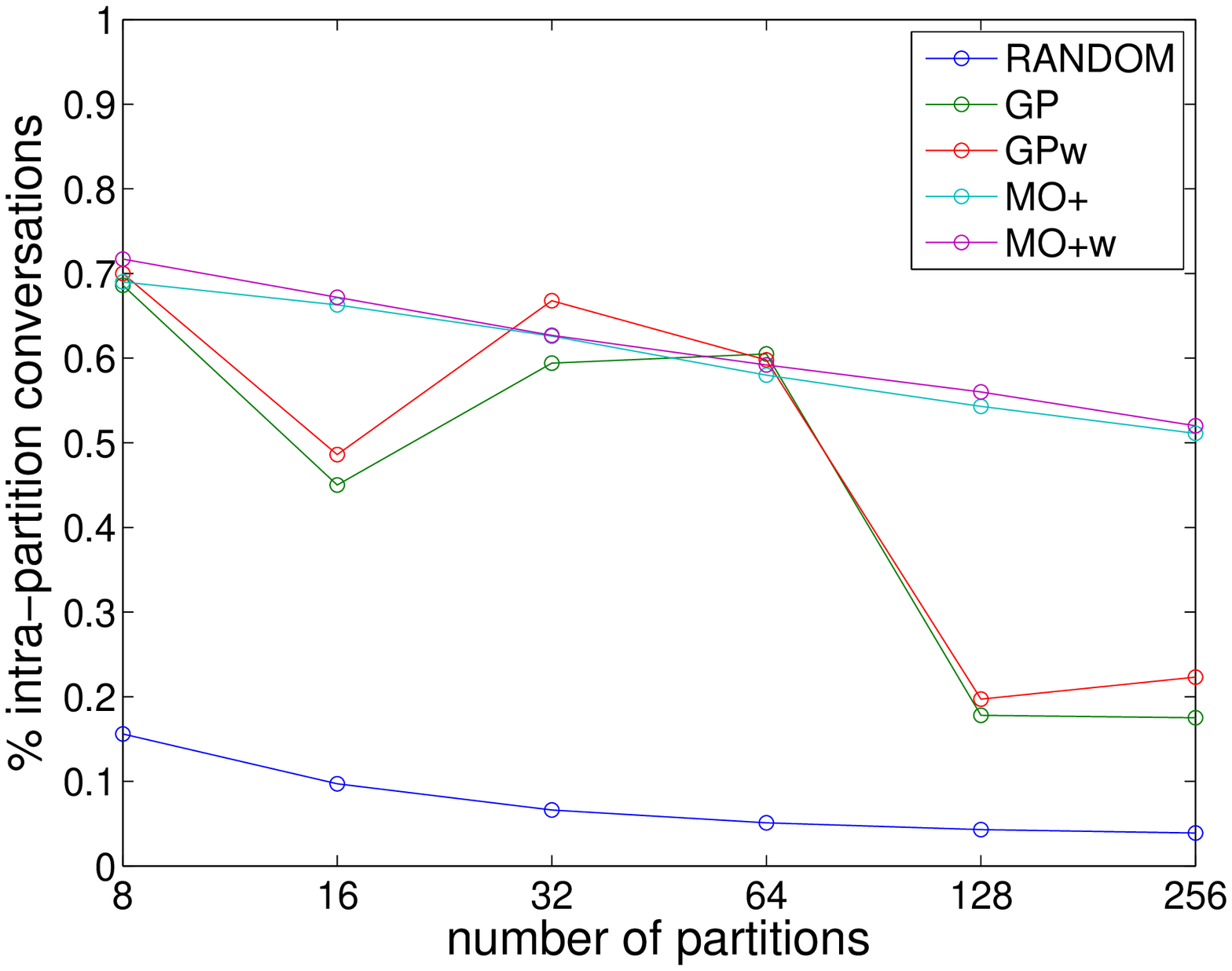}
\\
(a) & (b)
\end{tabular}
\caption{Partitioning Results with Traffic: \%of Intra-Partition Edges vs \# of Partitions (a) Twitter, (b) Conversations }
\label{fig:graph_part2}
\vspace{-3mm}
\end{figure*}

\textbf{Modularity Optimization Enhanced (MO+):}

 In order to obtain a predefined and fixed 
number of communities, each community of a fixed size, we modify the results obtained by 
a modularity optimizing algorithm. In order to obtain a predefined number of partitions 
of equal size we devised the algorithm MO+ \ref{algo1} that can be applied as a 
post-processing step to the results of a 
modularity optimization algorithm MO. At a high level, we start 
grouping the communities in a partition (with a predefined size) 
sequentially until the partition is full, when we move to next available partition. 
In the case the community does not fit, 
we apply MO to the subset graph recursively. This scheme is 
extremely simple and naive, but for our purpose, it suffices. 

\textbf{Random Partitioning:} As a baseline scheme, we also consider partitioning randomly. 
Random partitioning has few advantages - in addition to being very simple to implement,
randomization balances load optimally. Hence if we are only concerned 
with balancing load, then random partitioning will give the best results. However
such a scheme may not do well in reducing inter-partition traffic.
 

\vspace{-4mm}
\section{Experiments} 
\label{sec:exper}


\begin{figure*}[t]
\centering
\begin{tabular}[3]{ccc}
\includegraphics[width=2.2in]{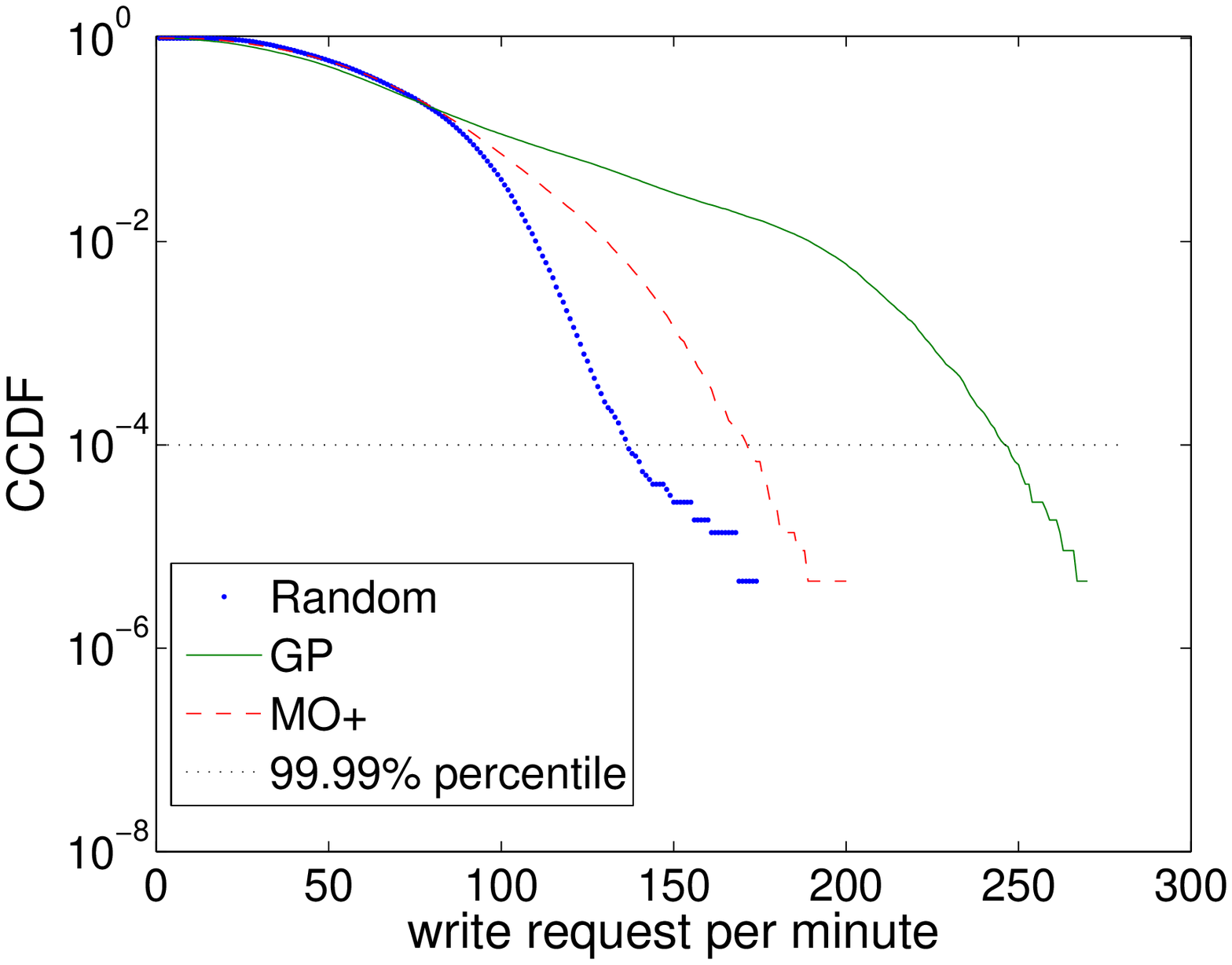}
&
\includegraphics[width=2.2in]{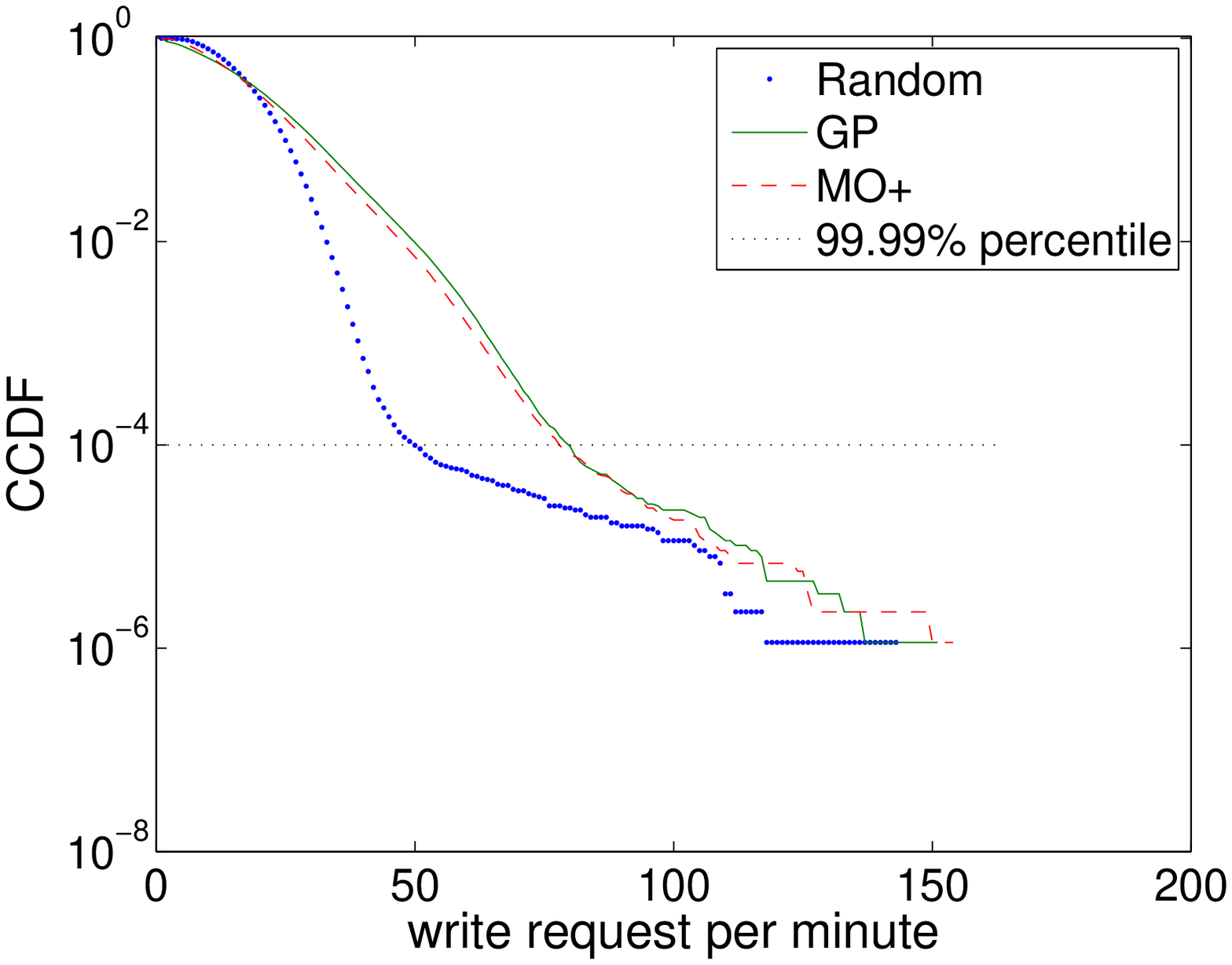}
&
\includegraphics[width=2.2in]{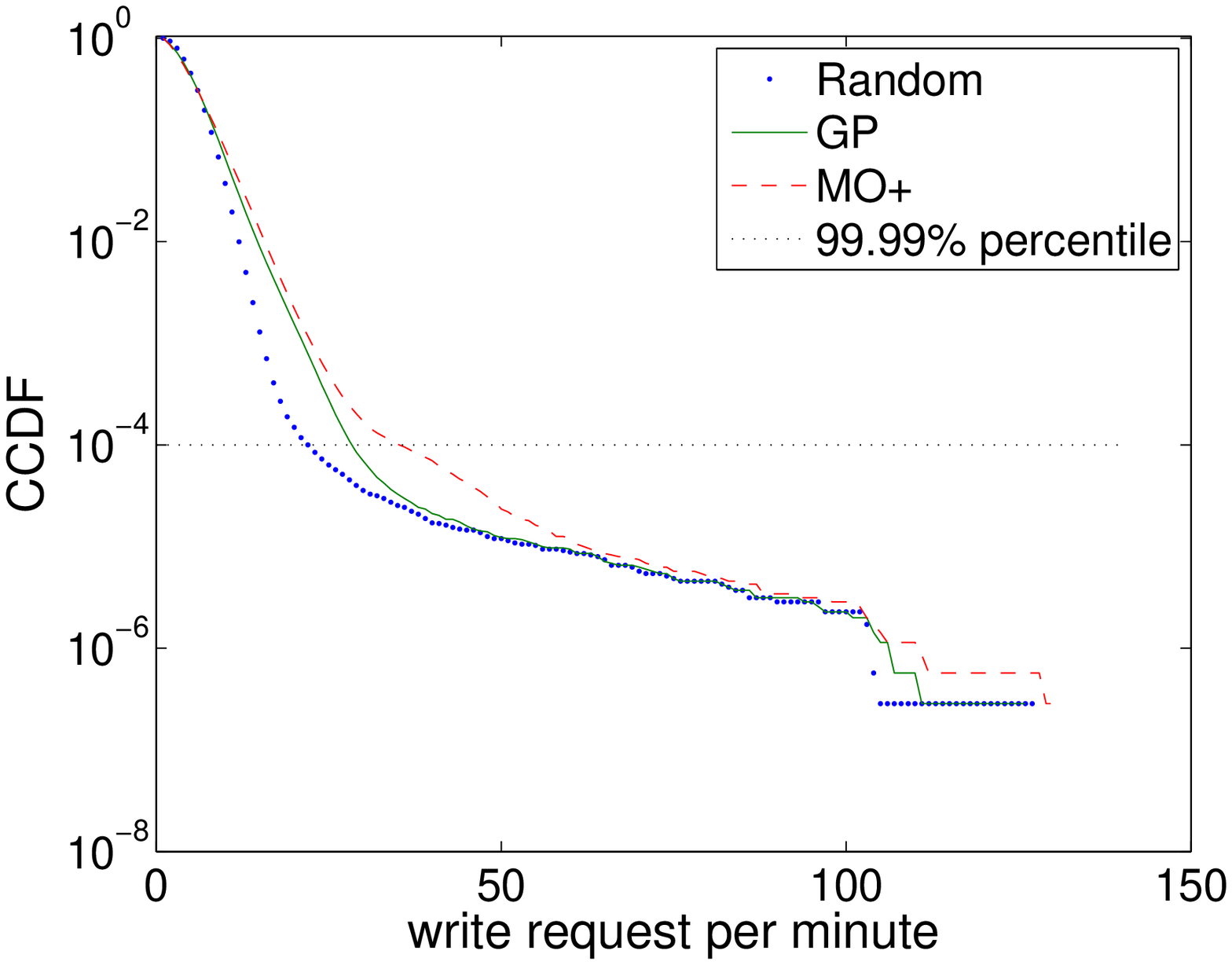}
\\
(a) & (b) & (c)
\end{tabular}
\caption{Distribution of Peak Load (a) 8 (b) 32 (c) 128 }
\label{fig:graph_load}
\vspace{-5mm}
\end{figure*}


The algorithms we try out include Random (R), Graph Partitioning (GP), 
Modularity Optimization+ (MO+) for Twitter and Orkut. GPw and MO+w stand for
the weighted versions of GP and MO+, where the weight can represent traffic on edges. 
We use the average of the number of tweets of exchanged between users;
in the case that edge weight is $0$, we set the weight to $1$ to avoid 
disconnecting the graph. We study the effect of weighted edges for 
Twitter only since we do not have access to the traffic of Orkut. 
\vspace{-2mm}
\subsection{Communication across Partitions}
We first evaluate how different partitioning schemes perform in terms of
the number of a) edges, b) messages and c) conversations that exist within partitions. 
Note that higher the \emph{intra-}partition, lower the network traffic between machines.


\textbf{Intra-Partition Edges:} Fig.~\ref{fig:graph_part} depicts 
the percentage of \emph{intra-}partition edges. 
First of all, this metric is low for the Random, 
and it decreases as $k^{-1}$ where $k$ is the number of partitions. 
The other schemes decrease 
at a much slower rate showing that they are able to maintain the 
structure of the underlying social network. 
In network traffic terms, Random with 256 partitions
results in a mere 0.4\% of intra-partition edges, while 
GP and MO+ produce 10\% to 30\% of intra-partition edges 
for Twitter. For Orkut, the reduction on network traffic 
is even higher - both GP and MO+ produce around 50\% intra-partition 
edges. GP and MO+ yield very similar results for Orkut and Twitter with 
small-size partitions. However, MO+ (and the weighted counterpart MO+w) 
seems to be more consistent across different number of partitions. 

\textbf{Intra-Partition Messages:} For the next set of results,
we use actual actual traffic, in the form of messages 
exchanged (1.7B messages, as reported in Sec.~\ref{subsec:prop}) and
run them against the partitions produced. Here we use directed edges. Note that this is qualitatively different from
merely counting the number of edges. 
Fig.~\ref{fig:graph_part2} (a) shows intra-partition messages and the results 
are qualitatively similar to results in Fig.~\ref{fig:graph_part}. 
Again, MO+ gives better partitions than GP when the number of partitions increase
and both of them perform better than Random.



\textbf{Intra-Partition Conversations:} 
Conversations link people who are likely to have a stronger social relationship, 
and therefore, they are more susceptible to consume each others 
user generated content \cite{Golder06,oh08}, e.g. videos, pictures, etc. 
Thus, it can be extrapolated that these links will generate additional 
traffic besides messages that we capture. 
The 265K conversational links also serve as a ground-truth. 
We expect the social network based partition schemes to be able to retrieve 
a majority of these strong links. Fig.~\ref{fig:graph_part2}(b) shows that 
the percentage of conversations is higher than the intra-partition edges; 
MO+ retains more than 50\% of these social links even when the number of 
partition is 256, stressing the importance of using schemes that are more 
aware of the underlying social structure.
\vspace{-2mm}
\subsection{Partitions: Balancing Load}

We showed that partitions based on the network structure(GP, MO+) can 
significantly reduce network traffic while compared to a 
Random. However, we also need to study the effect of these schemes 
on the load distribution across the machines that host these partitions. 
We extract the distribution of the arrival of tweets per minute from the 
trace we have.
\footnote{Note that a tweet is a write request, it has to be stored 
and distributed. Read operations are more common but require much less 
resources since systems are designed for faster reads. Common practice is 
that the QoS of a read request should be $< 200$ms, while for a write it can be 
an order of magnitude higher.} 
If there was no partitioning, a single machine should be 
dimensioned (CPU, memory, I/O) to deal with an average of 448 write requests/min. 
Most systems rely on worst-case provisioning. In order to perform worst-case analysis, we
extracted the 99.99\% percentile traffic rate that came to be 951 writes/min. 
In other words, without partitioning, this would be the peak traffic that a single machine
would have to handle.

In Fig.~\ref{fig:graph_load} we plot the CCDF of the write request (tweets arrival) 
per min across all partitions for different partition sizes (8,32,128 respectively). 
For 8 machines (Fig.~\ref{fig:graph_load} (a)), 
the 99.99\% percentile peak of traffic is 136, 171, 246 req/min for Random, MO+ and GP 
respectively. After partitioning the machines only need to have resources to deal 
with this reduced load peak. For the partition in 
128 machines (Fig.~\ref{fig:graph_load} (c)) the load peak is 
21,35,28 req/min for Random, MO+ and GP.

The picture that emerges from our experiments is as follows. 
Schemes based on modularity optimization are good for reducing network traffic,
and are competitive with Random in terms of load balancing across partitions.
Hence while the tradeoff between reducing bandwidth costs and balancing load
is still present, we find that MO based schemes are promising. 
We interpret these results in light of the arguments we made in Sec~\ref{subsec:prop},
more specifically the existence of social conversations, and the 
community structure.

%
 

\vspace{-4mm}

\section{Discussion}
\label{sec:discuss}
Online social networks are increasing at a very high rate, putting a strain on existing
infrastructure and demanding new architectural solutions. As such systems get pushed
to highly distributed cloud infrastructure, the problem of finding intelligent solutions
to scaling issues is further exacerbated.

In this paper we explored the problem of partition the users' space of different OSN using
different schemes. By using data from real OSN -- Twitter and Orkut -- 
we measured the impact of different partition schemes on network traffic 
and peak load. We found that traditional graph partitioning techniques 
can effectively be applied to OSNs to reduce the network overhead that 
a typical off-the-shelf random partition would generate while maintaining an 
acceptable peak load distribution. We also show that modularity optimization algorithms are
even better suited to the social network partition problem given that they are based on 
finding `community' structures in the network. Such a scheme, however, presents the problem
of producing arbitrary number of communities and we propose a post-processing algorithm 
MO+ to handle this issue.




{\footnotesize 
\bibliographystyle{abbrv}
\bibliography{paper}
}

\end{document}